\documentclass[conference]{IEEEconf}
\IEEEoverridecommandlockouts

\usepackage{cite}
\usepackage{amsmath,amssymb,amsfonts}
\usepackage{graphicx}
\usepackage{textcomp}
\usepackage[table]{xcolor}
\usepackage{caption}
\usepackage{lettrine}
\usepackage[
    hidelinks,
    bookmarks=false,
    unicode=false,
    pdfencoding=auto
]{hyperref}
\usepackage{multirow}
\usepackage{booktabs}
\usepackage{makecell}
\usepackage{subfig}

\usepackage{lipsum}
\usepackage[ruled,vlined]{algorithm2e}

\begin{document}

\title{\LARGE \bf Separation Assurance between Heterogeneous Fleets of Small Unmanned Aerial Systems via Multi-Agent Reinforcement Learning}

\author{
Iman Sharifi$^{1}$,
Hyeong Tae Kim$^{2}$,
Maheed Hatem Ahmed$^{2}$,
Mahsa Ghasemi$^{2}$,
and Peng Wei$^{1}$%
\thanks{$^{1}$I. Sharifi and P. Wei are with the Department of Mechanical and Aerospace Engineering, George Washington University, Washington, DC, USA.
        {\tt\small \{i.sharifi,pwei\}@gwu.edu}}%
\thanks{$^{2}$H. T. Kim, M. H. Ahmed, and M. Ghasemi are with the Department of Electrical and Computer Engineering, Purdue University, West Lafayette, IN, USA.
        {\tt\small \{kim4741,ahmed237,mahsa\}@purdue.edu}}%
}
\maketitle

\begin{abstract}
In the envisioned future dense urban airspace, multiple companies will operate heterogeneous fleets of small unmanned aerial systems (sUASs), where each fleet includes several homogeneous aircraft with identical policies and configurations, e.g., equipage, sensing, and communication ranges, making tactical deconfliction highly complex for the aircraft. This paper aims to address two core questions: (1) Can tactical deconfliction policies converge or reach an equilibrium to ensure a conflict-free airspace when companies operate heterogeneous fleets of homogeneous aircraft? (2) If so, will the converged policies discriminate against companies operating sUASs with weaker configurations? We investigate a multi-agent reinforcement learning paradigm in which homogeneous aircraft within heterogeneous fleets operate concurrently to perform package delivery missions over Dallas, Texas, USA. An attention-enhanced Proximal Policy Optimization-based Advantage Actor-Critic (PPOA2C) framework is employed to resolve intra- and inter-fleet conflicts, with each fleet independently training its own policy while preserving privacy. Experimental results show that two fleets with distinct, shared PPOA2C policies can reach an equilibrium to maintain safe separation. While two PPOA2C policies outperform two strong rule-based baselines in terms of conflict resolution, a PPOA2C policy exhibits safer interaction with a rule-based policy, indicating adaptive capabilities of PPOA2C policies. Furthermore, we conducted extensive policy-configuration evaluations, which reveal that equilibria between similar policy types tend to favor fleets with stronger configurations. Even under similar configurations but different policy types, the equilibrium favors one of the heterogeneous policies, underscoring the need for fairness-aware conflict management in heterogeneous sUAS operations.
\end{abstract}

\section{Introduction}

Unmanned aerial systems (UASs) are increasingly being adopted across a wide range of commercial and industrial domains, including infrastructure inspection, aerial surveying, environmental monitoring, and package delivery~\cite{ariante2025unmanned}. Specifically, small unmanned aerial systems (sUASs) represent a rapidly expanding class of low-altitude UASs whose compact design, low cost, and operational flexibility make them particularly well-suited for high-frequency, localized missions~\cite{beard2012small}. Companies such as Google Wing, Amazon Prime Air, and Zipline are increasingly deploying fleets of sUASs to perform time-sensitive and spatially distributed delivery tasks. As these operations scale, UAS Traffic Management (UTM) service providers are expected to accommodate sUAS flight operations in low-altitude urban airspace, particularly in densely populated areas. In such shared airspace, each company often deploys vehicles with distinct aircraft performance, sensing and communication ranges, and proprietary control policies, making tactical deconfliction among sUASs highly challenging. Tactical deconfliction—also known as conflict resolution—requires rapid, reactive decision-making based on local and dynamic state information to ensure safe separation~\cite{wang2022review}.

To address these inherent safety challenges in multi-agent UASs, various methods have been proposed~\cite{hunter2019service, cai2019integrated, ong2017markov}, among which multi-agent reinforcement learning (MARL) approaches have demonstrated superior reliability in maintaining separation assurance under highly dense traffic scenarios~\cite{wang2022review}. In an MARL framework, each agent typically optimizes its policy in either a fully centralized~\cite{brittain2019autonomous, brittain2021one, guo2021safety, brittain2024improving} or fully decentralized manner~\cite{alvarez2019acas, brittain2022scalable}. However, in real-world urban aerial traffic, a hybrid combination of centralized and decentralized learning often provides a more practical solution to the challenges of a high-density shared airspace with multiple operator companies. While aircraft (agents) within a company typically employ identical deconfliction logic or policies, companies do not share their policies with other stakeholders due to privacy and safety concerns. To this end, a hybrid decentralized-centralized approach not only allows homogeneous agents within a company to share experiences for improved training efficiency but also enables heterogeneous fleets between different companies to preserve policy confidentiality, enhancing overall safety.

In this paper, we investigate an MARL paradigm within a two-fleet, high-density scenario in which heterogeneous fleets of sUASs, each with multiple homogeneous aircraft, operate concurrently in a complex urban airspace to perform package delivery missions. Each fleet, representing a specific company such as Google Wing or Amazon Prime Air, consists of autonomous agents with distinct aircraft capabilities—such as maximum speed, acceleration, and sensor or communication range (e.g., one company uses radar detection for intruders, while the other employs Remote ID~\cite{faa2020remoteid} for vehicle-to-vehicle information sharing). Each fleet independently trains its own MARL policy using an attention-based PPO-driven Advantage Actor-Critic (PPOA2C) framework~\cite{brittain2019autonomous, brittain2022scalable} to learn optimal tactical deconfliction strategies for both homogeneous and heterogeneous agents. While the architectural backbone of the method remains identical, the policies are trained separately, resulting in heterogeneous behaviors that reflect both differing vehicle dynamics and independent optimization processes. This setup mirrors realistic constraints found in competitive and proprietary operational environments.

In particular, we seek to answer the following essential questions: (1) Will two fleets with proprietary PPOA2C policies lead to converged tactical deconfliction policies or reach an equilibrium for a conflict-free airspace, given that companies operate heterogeneous aircraft types with distinct sensing and communication ranges? (2) If so, will the equilibrium discriminate against a company operating sUASs with weaker performance, inferior equipage, or shorter sensing and communication ranges? We aim to investigate both operational safety and fairness challenges across heterogeneous sUAS fleets. 

The main contributions of this paper are as follows:
\begin{itemize}
    \item We employ PPOA2C in a dense air traffic scenario including two heterogeneous fleets of sUASs, striving to ensure safe separation given the heterogeneity in both distributed policies and agents' physical configurations. We demonstrate that heterogeneous MARL policies for mixed aircraft types can reach an equilibrium. 
    \item We show that the PPOA2C method outperforms a strong rule-based method in a dense scenario over Dallas, Texas. Moreover, PPOA2C not only cooperates safely with another PPOA2C policy but also learns to interact safely with a rule-based method.
    \item Through extensive policy-configuration evaluations, we show that the equilibrium with converged policies tends to favor aircraft with stronger configurations. Furthermore, even with identical configurations, training between two different policy types can exhibit discrimination against one of them.
\end{itemize}

The remainder of this paper is organized as follows. Section~\ref{sec:related_work} reviews the related literature. Section~\ref{sec:methodology} presents the problem formulation and technical approach. Section~\ref{sec:experimental-results} details the experimental setup and analyzes the results. Finally, Section~\ref{sec:conclusion} concludes the paper.

\section{Related Work}\label{sec:related_work}

Two types of tactical deconfliction strategies based on MARL have been proposed to improve separation assurance in air traffic control~\cite{wang2022review}: centralized training with decentralized execution (CTDE) and decentralized training and execution (DTE). The CTDE method leverages either global or local information from agents to train a centralized unit, such as a global critic or a shared actor-critic neural network, which governs agents' behaviors during deconfliction in high-density en-route airspace. Brittain and Wei~\cite{brittain2019autonomous} introduced a multi-agent actor–critic framework that combines the advantage actor-critic (A2C) algorithm with elements of proximal policy optimization (PPO) under a centralized training, decentralized execution paradigm. To accommodate a variable number of agents, Brittain and Wei~\cite{brittain2021one} proposed a long short-term memory (LSTM) network, enabling each agent to process a flexible number of intruder inputs. More recent approaches address the challenge of handling varying numbers of intruders by employing graph convolutional neural networks~\cite{isufaj2022multi} and attention mechanisms~\cite{brittain2021autonomous, brittain2024improving}. Groot et al.~\cite{groot2025comparing} compared three different attention mechanisms—scaled dot-product, additive, and context-aware attention—integrated with the soft actor-critic (SAC) algorithm to manage separations in high-density traffic scenarios. Other centralized reinforcement learning approaches integrate classical models—such as the 3D reciprocal velocity obstacle principle~\cite{zhong20253d} and the solution space diagram method~\cite{zhao2021physics}—to enhance the quality of the observations provided to the reinforcement learning agent. In contrast, the DTE framework assumes that each agent’s training and policy are independent of those of other agents. 

While the CTDE framework may suffice for homogeneous fleets, heterogeneous UAS operations demand more flexible architectures, which can be supported by DTE. In a fully decentralized setting, Brittain and Wei~\cite{brittain2022scalable} presented a distributed deep reinforcement learning approach using an off-policy SAC algorithm enhanced with attention networks. Our method combines the advantages of centralized learning with the scalability of decentralization: within a single sUAS fleet, agents share parameters and data to improve sample efficiency, while training remains decentralized across different fleets to ensure both policy flexibility and privacy.

\section{Problem Formulation and Methodology} \label{sec:methodology}
\subsection{\textbf{Tactical Deconfliction Strategy}}

In future airspace operations, a combination of CTDE and DTE strategies is more realistic than fully centralized or fully decentralized approaches, as it accommodates different policies for each fleet while allowing policy sharing among agents within the same fleet. The primary objective of such mixed frameworks is to resolve both intra- and inter-fleet conflicts through an intelligent autonomous framework in which each fleet’s control policy is trained independently and remains private from other entities. This hybrid strategy distributes a shared policy among homogeneous agents while distinguishing between heterogeneous fleets’ policies. It aims to ensure safe separation among both homogeneous and heterogeneous agents in dense aerial traffic environments where multiple companies simultaneously utilize shared airspace. Although a common MARL framework is adopted across fleets, each fleet’s agents have distinct configurations and train independently using their own collected experiences. Consequently, the learning processes and policy parameters remain fully independent, and information sharing occurs only through publicly available data. Through MARL, agents can adapt to dynamic environments and improve over time by interacting not only with homogeneous teammates but also with heterogeneous agents from other fleets.

\subsection{\textbf{Multi-Agent PPO-driven Advantage Actor-Critic}}

To jointly train homogeneous agents within heterogeneous fleets, we employ an advantage actor-critic (A2C) framework~\cite{mnih2016asynchronous} combined with a loss function derived from proximal policy optimization (PPO)~\cite{schulman2017proximal, yu2022surprising}, similar to~\cite{brittain2021autonomous}. A2C is a policy-gradient method that often uses a unified neural network architecture to estimate both the policy (actor) and the value (critic) functions. PPO complements A2C by introducing a clipping-based loss function that stabilizes learning, ensuring that policy updates remain within a controlled range of the previous policy. We refer to the combination of PPO and A2C as PPOA2C, which enables efficient exploration of the action space and allows agents to refine their strategies by focusing on rewarding behaviors, ultimately improving robustness and convergence performance within the decentralized multi-agent setting.

To enhance the scalability and adaptability of the PPOA2C framework in high-density airspace environments, we incorporate a multiplicative attention mechanism~\cite{brittain2024improving, groot2025comparing} into the A2C neural network architecture to capture high-level relationships among states. This extension enables each aircraft agent to selectively attend to the most relevant intruders—an essential capability in scenarios where the number of nearby intruders varies dynamically. The attention module encodes a variable number of neighboring agents into a fixed-length context vector, which is subsequently passed to the policy and value networks.
After computing action probabilities and the state value using the A2C network, the optimization process minimizes two primary loss components, including the policy loss:
\begin{align*}
\mathcal{L}_\pi = & \ \mathbb{E}_t\left[\min\left(\zeta_t(\theta) \cdot A_t,\; \text{clip}(\zeta_t(\theta), 1 - \epsilon, 1 + \epsilon) \cdot A_t\right)\right] \nonumber \\
& - \beta \ H(\pi(s_t)),
\end{align*}
and the value loss: \(\mathcal{L}_v = A_t^2\). The advantage function $A_t$ quantifies the relative quality of an action compared to the expected behavior under the current policy. To compute $A_t$ more accurately and robustly, we employ the Generalized Advantage Estimation (GAE) technique~\cite{schulman2015high}. The term $\zeta_t(\theta)$ is the likelihood ratio between the new and old policies. The parameter $\epsilon$ defines the clipping range that restricts the magnitude of policy updates. The entropy regularization term $\beta \cdot H(\pi(s_t))$ encourages exploration by penalizing premature convergence to deterministic policies, where $H(\pi(s_t))$ denotes the entropy of the current policy distribution and $\beta$ controls its influence during training. 
\begin{figure}
    \centering
    \vspace{2mm}
    \includegraphics[width=0.9\linewidth, trim={3mm 3mm 3mm 0mm}, clip]{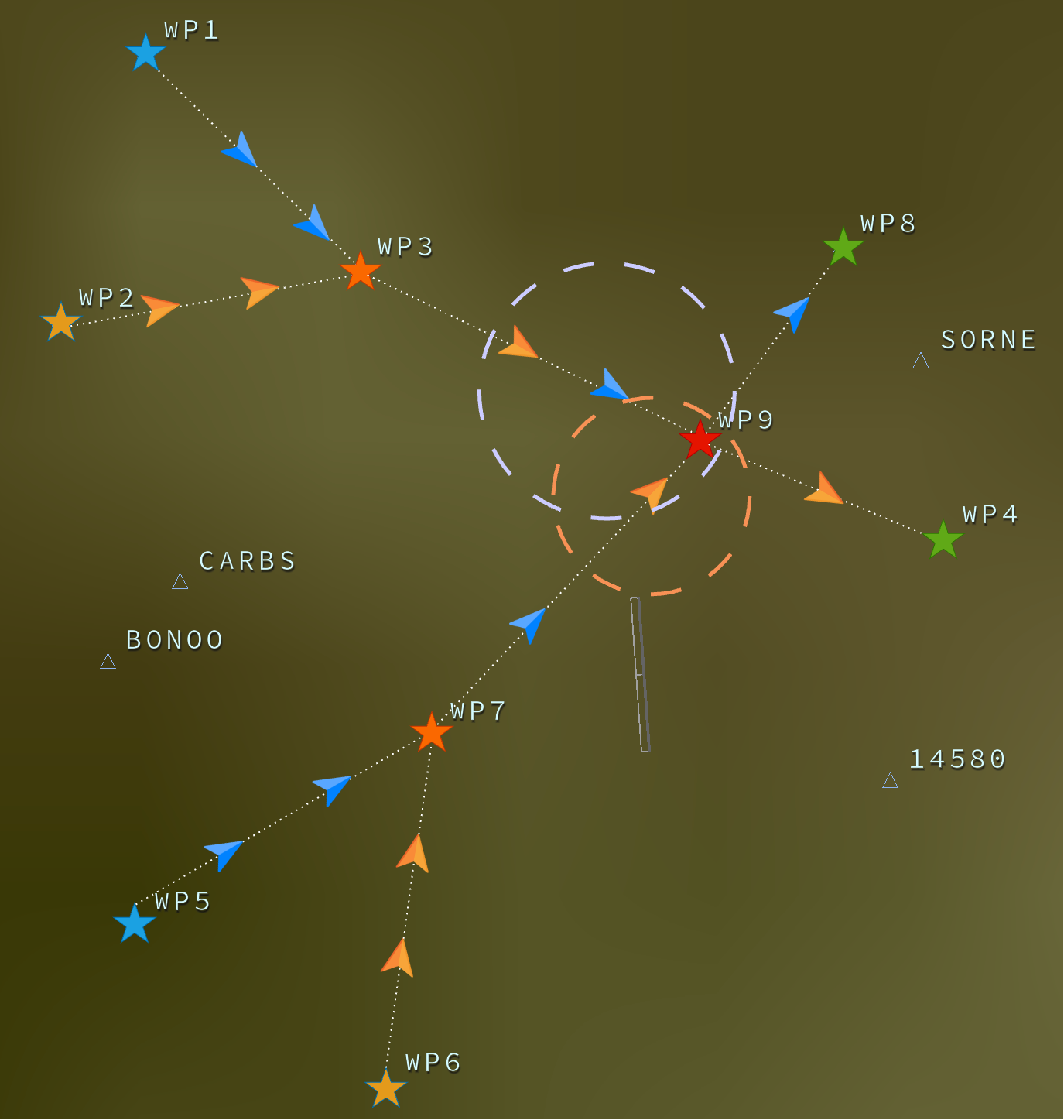}
    \caption{Use-case scenario based on Frisco, a suburb area in Dallas, Texas, simulated in BlueSky. Merging points \texttt{WP3} and \texttt{WP7} and the Intersection \texttt{WP9} correspond to \texttt{M1}, \texttt{M2}, and \texttt{IN} bottlenecks, respectively.}
    \label{fig:scenario}
    \vspace{-4mm}
\end{figure}
\subsection{\textbf{MARL Components}}
The core MARL components are outlined as follows:

\subsubsection{\textbf{State Space}} 
In reinforcement learning, the state space represents the set of all possible environmental configurations that an agent can observe at a given time. Here, we assume that aircraft state and dynamic information are openly shared among agents. This assumption reflects realistic operational considerations, as the Federal Aviation Administration (FAA) mandates Remote ID~\cite{faa2020remoteid} to broadcast essential information for all participating sUASs. According to FAA standards, all agents broadcast basic information, including aircraft ID, location, altitude, and speed, to neighboring aircraft and the nearest ground control station. Thus, we assume each agent has access to the basic information of nearby agents.

The states for each agent are divided into two components: the ownship state and the intruders state, defined as:
\begin{equation*}
s_t^o = [ d^{(o)},\ v^{(o)},\ \theta^{(o)},\ v'^{(o)}], \;\;
s_t^i = [ d_o^{(i)},\ v^{(i)},\ \theta^{(i)},\ v'^{(i)}],
\label{eq:ownship_intruder_state}
\end{equation*}
where $s_t^o \in \mathbb{R}^{|s^o|}$ denotes the ownship state, consisting of the distance to the next waypoint $d^{(o)}$, current speed $v^{(o)}$, heading $\theta^{(o)}$, and acceleration $v'^{(o)}$. Except $d_o^{(i)}$ which is the distance between ownship $o$ and intruder $i$, the intruder state $s_t^i \in \mathbb{R}^{|s^i|}$ mirrors the ownship state structure. Intruder is an aircraft within a certain range around the ownship. To avoid complexity in the decision-making, we only consider front intruders, as aircraft arriving at the next bottleneck earlier than the ownship, since they have the most direct influence on the ownship’s immediate tactical decisions. Following~\cite{chen2024integrated}, this approach improves both computational and operational efficiency. After extracting the states, each component of $s_t^o$ and $s_t^i$ undergoes a normalization step before being fed into the PPOA2C neural network.

\subsubsection{\textbf{Action Space}} 
In this study, the action space is defined as the set of discrete speed adjustments that an aircraft can apply at each decision step. The agent can choose to \textit{decelerate}, \textit{hold} its current speed, or \textit{accelerate}:
\(\mathbb{A} = \{-\Delta v,\; 0,\; +\Delta v\}\),
where $\Delta v$ is the speed adjustment magnitude specific to each fleet of agents and varies depending on their configurations. After selecting an action $a_t^o \in \mathbb{A}$, it is applied to the ownship’s current speed and transmitted to the simulation environment. 

\subsubsection{\textbf{Reward Function}} 
In reinforcement learning, the reward function provides a scalar feedback signal that reflects the desirability of the state–action pair executed by the agent. In our framework, the total reward $r_t^o$ for each ownship agent at each time step is composed of five distinct components:
\begin{equation*}
    r_t^o = R_{\text{LoS}} + R_{\text{V}} + R_{\text{A}} + R_{\text{M}} + R_{\text{T}},
    \label{eq:reward_components}
\end{equation*}
where $R_{\text{LoS}}$, known as the loss of separation (LoS) reward, penalizes unsafe separation; $R_{\text{V}}$ penalizes undesirable velocities; $R_{\text{A}}$ penalizes undesirable flight behaviors such as abrupt speed changes; $R_{\text{M}}$ incentivizes task completion; and $R_{\text{T}}$ encourages time efficiency.

Since maintaining safe separation is the central objective of this study, $R_{\text{LoS}}$ constitutes the primary term in the reward function. It is defined as:
\begin{equation*}
R_{\text{LoS}} =
\begin{cases}
    -1, & \text{if } d_o^{(i)} < d_{\text{NMAC}}, \\
    \alpha\left(-1 + \frac{d_o^{(i)} - d_{\text{NMAC}}}{d_{\text{LoWC}} - d_{\text{NMAC}}}\right), & \text{if } d_{\text{NMAC}} \leq d_o^{(i)} \leq d_{\text{LoWC}}, \\
    0, & \text{otherwise},
\end{cases}
\end{equation*}
where $d_o^{(i)}$ denotes the distance between the ownship and the $i$-th intruder, $d_{\text{NMAC}}$ is the near mid-air collision (NMAC) threshold, and $d_{\text{LoWC}}$ is the loss of well clear (LoWC) threshold. If the separation distance falls below $d_{\text{NMAC}}$, a severe penalty of $-1$ is imposed, immediately removing the ownship and the corresponding intruder agent from the simulation. If the distance lies between $d_{\text{NMAC}}$ and $d_{\text{LoWC}}$, the penalty is linearly scaled with distance using the coefficient $\alpha \in [0,1]$.

To ensure that the agents’ speeds remain within a specified range, another reward component penalizes violations of the speed limits, as defined below:
\begin{equation*}
R_{\text{V}}
=-\psi_1^v \cdot \mathbb{I}\!\ [v^{(o)} < v_{\text{min}}^{(o)} + \eta_1^v] - \psi_2^a \cdot \mathbb{I}\!\ [v^{(o)} > v_{\text{max}}^{(o)}-\eta_2^v], 
\end{equation*}
where $\mathbb{I}[\cdot]$ equals $1$ if the condition holds, and $0$ otherwise. $v_{\text{min}}^{(o)}$ and $v_{\text{max}}^{(o)}$ are the nominal minimum and maximum speeds of the ownship. The parameters $\eta_1^v$ and $\eta_2^v$ are offset values that prevent agents from reaching the exact minimum and maximum velocities, respectively. The hyperparameters $\psi_1^v$ and $\psi_2^v$ assign small penalties when these limits are approached. These constraints prevent agents from becoming stuck in the environment and encourage energy efficiency. 

In real-world operations, frequent or abrupt speed adjustments in urban environments can destabilize aircraft—particularly those carrying payloads—and increase risks for nearby traffic, thereby compromising safety. To discourage such undesirable behaviors, we introduce an action penalty term $R_{\text{A}}$, defined as:
\begin{equation*}
R_{\text{A}}
=-\psi_1^a \cdot \mathbb{I}\!\ [a_t^{(o)} \neq a_{t-1}^{(o)}] -\psi_2^a \cdot \mathbb{I}\!\ [a_t^{(o)} \neq \texttt{HOLD}], 
\end{equation*}
where $a_t^{(o)}$ and $a_{t-1}^{(o)}$ denote the current and previous actions of the ownship agent, respectively. The hyperparameters $\psi_1^a$ and $\psi_2^a$ control the penalties for frequent speed changes and deviations from steady flight. This formulation encourages agents to maintain smoother and more consistent speed profiles, thereby promoting both energy conservation and operational safety.

We also incorporate a reward component to promote mission accomplishment, represented by: \(
R_{\text{M}}
=\psi_m \cdot \mathbb{I}\!\ [d_f^{(o)} < \eta_m] 
\), which provides a bonus for reaching the final destination.
$d_f^{(o)}$ denotes the distance between the ownship and its final waypoint, and $\eta_m$ is a predefined distance threshold. If the aircraft reaches within a specified proximity to its destination, it receives a terminal reward of $\psi_m$ and is withdrawn from the simulation; otherwise, it receives no penalty.

To encourage timely mission completion, a small negative penalty $-\psi_t$ is applied at every time step. This component incentivizes efficient traversal toward the goal, as:
\(
R_{\text{T}}
=-\psi_t \cdot \mathbb{I}\!\ [t < T] -\mathbb{I}\!\ [t \geq T]
\), where $T$ is a time threshold after which the agent can no longer operate in the environment and is terminated from the simulation. Then, the agent receives a penalty of $-1$.
Overall, the reward function prioritizes safety, efficiency, and smoothness in flight operations. 

\subsection{\textbf{Training Approach}} 

After collecting all state information for each agent within a fleet, the corresponding PPOA2C network architecture first processes $s_t^o$ and $s_t^i$ through separate $64$-unit fully connected layers, followed by a $64 \times 64$ attention module to capture interaction features. The resulting representation is further transformed by two $128$-unit fully connected layers before branching into the actor and critic heads to generate the action probability distribution and the state-value estimate. Based on this probability distribution, an action $a_t^o$ is sampled for each ownship agent; a reward $r_t^o$ is then assigned accordingly, and the tuple $(s_t^o, s_t^i, a_t^o, r_t^o)$ is stored in the fleet replay buffer. This process continues until the episode terminates.

After each episode, discounted returns and advantage estimates are computed from the collected trajectories and appended to the corresponding samples. Training then proceeds by evaluating the PPOA2C loss over a batch of $512$ samples. The network parameters are updated using the Adam optimizer, and this optimization procedure is repeated for $8$ epochs per episode for each fleet. Once the policy and value networks have been optimized, the resulting policy is shared among all homogeneous agents within that fleet. This process is executed independently for each fleet, after which the learned policies are deployed to their respective agents, thereby preserving policy coherence within fleets while maintaining heterogeneity across fleets.

The learning rate, entropy coefficient $\beta$, and clipping threshold $\epsilon$ are $10^{-4}$, $10^{-3}$, and $0.2$, respectively.
The separation thresholds $d_{\text{NMAC}}$ and $d_{\text{LoWC}}$ are set to 
$100$\,m and $500$\,m, respectively, and the goal tolerance $\eta_m$ is $50$\,m. 
The simulation time step $\Delta t$ and mission horizon $T$ are 
$3$\,s and $18$\,min, respectively. 
The speed offsets $\eta_1^v$ and $\eta_2^v$ are 
$5.14$\,m/s and $2.57$\,m/s, respectively. 
The reward-shaping coefficients $\psi_1^v$ and $\psi_2^v$ are 
$10^{-3}$ and $10^{-4}$, respectively, while 
$\psi_1^a$ and $\psi_2^a$ are 
$10^{-5}$ and $10^{-4}$, respectively. 
The remaining coefficients are $\alpha = 0.1$, 
$\psi_t = 10^{-4}$, and $\psi_m = 0.1$.

\section{Experimental Results}\label{sec:experimental-results}
\subsection{\textbf{Simulation Setup}}

To model and evaluate the performance of our tactical deconfliction framework, we employ BlueSky~\cite{hoekstra2016bluesky}, an open-source, fast-time air traffic simulator widely recognized in the aviation research community for its versatility and scalability. 

\subsubsection{\textbf{Use-case Scenario}}
To train MARL policies in a realistic and structured environment, we design a custom scenario based on the airspace over Frisco, a suburban area in Dallas, Texas, as illustrated in Fig.~\ref{fig:scenario}. The scenario consists of four fixed routes representing typical drone delivery paths:

\begin{itemize}
    \item Route~I: \texttt{WP1} $\rightarrow$ \texttt{WP3} $\rightarrow$ \texttt{WP9} $\rightarrow$ \texttt{WP4}
    \item Route~II: \texttt{WP2} $\rightarrow$ \texttt{WP3} $\rightarrow$ \texttt{WP9} $\rightarrow$ \texttt{WP4}
    \item Route~III: \texttt{WP5} $\rightarrow$ \texttt{WP7} $\rightarrow$ \texttt{WP9} $\rightarrow$ \texttt{WP8}
    \item Route~IV: \texttt{WP6} $\rightarrow$ \texttt{WP7} $\rightarrow$ \texttt{WP9} $\rightarrow$ \texttt{WP8}
\end{itemize}

The origin points of these routes (marked by blue and orange stars) are selected approximately based on the locations of major commercial hubs such as Walmart and Walgreens. The destination points (green stars) correspond to residential areas where delivery demand is expected to be high during the day. The routes are designed to have approximately the same total length of $10.33$~km for both companies. To simulate a realistic and challenging operational environment, the routes are deliberately configured to create bottlenecks—specifically at \texttt{WP3} and \texttt{WP7} as merging waypoints and \texttt{WP9} as an intersection waypoint—which reflect common congestion points in low-altitude airspace networks.

In this scenario, we simulate a total of $20$ agents, with $10$ agents assigned to each company. Company~A (Co.~A) agents operate on Route~I and Route~III (orange stars in Fig.~\ref{fig:scenario}), while Company~B (Co.~B) agents operate on Route~II and Route~IV (blue stars in Fig.~\ref{fig:scenario}). This configuration results in five agents per route. Agent spawn times (in seconds) are defined as $35 + 5k$, where $k$ is randomly selected from $\{0, 1, \dots, 10\}$ to introduce temporal variability. This design not only generates numerous conflict situations at bottlenecks but also provides sufficient temporal spacing for agents to make informed and strategic decisions. Conflicts are expected to arise at merging points \texttt{WP3} and \texttt{WP7}, where agents from both companies converge. However, the most complex and congested interactions occur at the intersection point \texttt{WP9}. This setup reflects realistic operational conditions in shared urban airspace, where agents must resolve both intra- and inter-company conflicts.

\subsubsection{\textbf{sUAS Configurations}}
To model heterogeneity, we consider distinct configurations for each fleet of agents, including different speed limits, acceleration ranges, and sensory capabilities, reflecting the diversity of hardware and onboard systems across drone operators. In general, we consider two configurations: \texttt{X} and \texttt{Y}, where configuration \texttt{X} possesses stronger capabilities than configuration \texttt{Y}. The speed limits and acceleration ranges for configurations \texttt{X} and \texttt{Y} are selected based on the performance specifications of the Google Wing Hummingbird drone and the Amazon MK30 drone, respectively. The sensory ranges are chosen to align with current technological standards for drone communication via Remote ID~\cite{faa2020remoteid} or radar detection, ensuring that agents can detect and respond to nearby intruders while maintaining distinct sensing ranges. Configurations \texttt{X} (strong) and \texttt{Y} (weak) have speed ranges $[0, 44.88]$ and $[0, 30.12]$ m/s, acceleration sets $\{-1.71, 0, 1.71\}$ and $\{-1.02, 0, 1.02\}$ m/s$^2$, and sensing ranges of $1000$ and $750$ m, respectively.

The ultimate goal is to train all agents with both homogeneous and heterogeneous configurations such that they learn to minimize the occurrence of NMACs during operation while successfully completing their mission. The desired outcome is a cooperative and adaptive system in which all agents fly in the shared airspace safely and efficiently despite differences in control policies, aircraft performance, and sensor capabilities. In particular, agents are expected to coordinate not only with teammates trained under the same policy but also with agents operating under independently trained policies from competing companies.

\begin{figure}
    \centering
    \includegraphics[width=\linewidth]{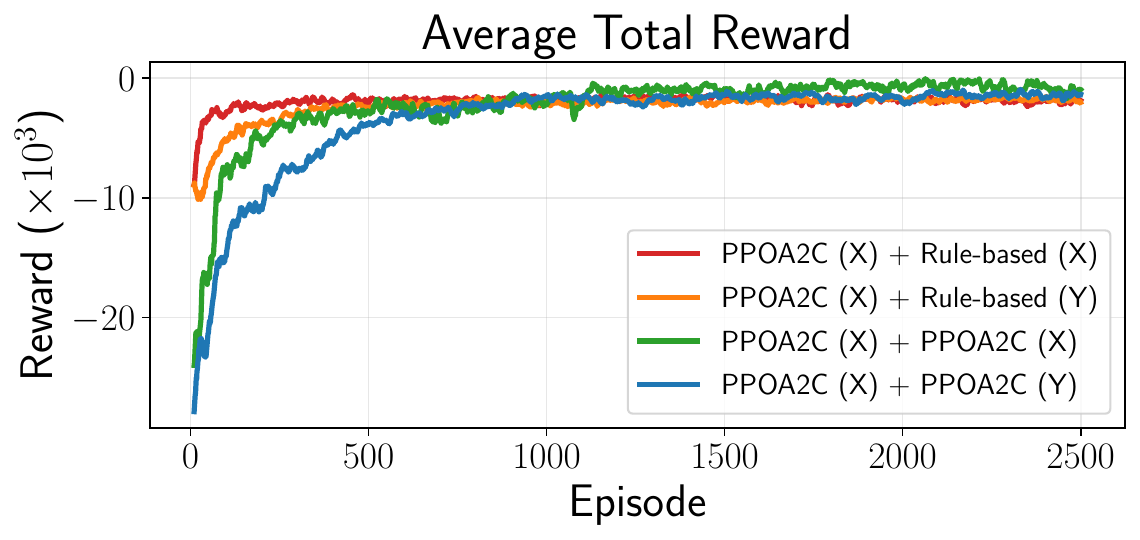}
    \caption{Average total rewards gained by both Co.~A and Co.~B agents with various policy-configuration models.}
    \label{fig:avg-tot-rewards}
    \vspace{-4mm}
\end{figure}


\begin{figure*}[h]
    \centering
    \subfloat[Average Reward]{\includegraphics[width=0.48\textwidth]{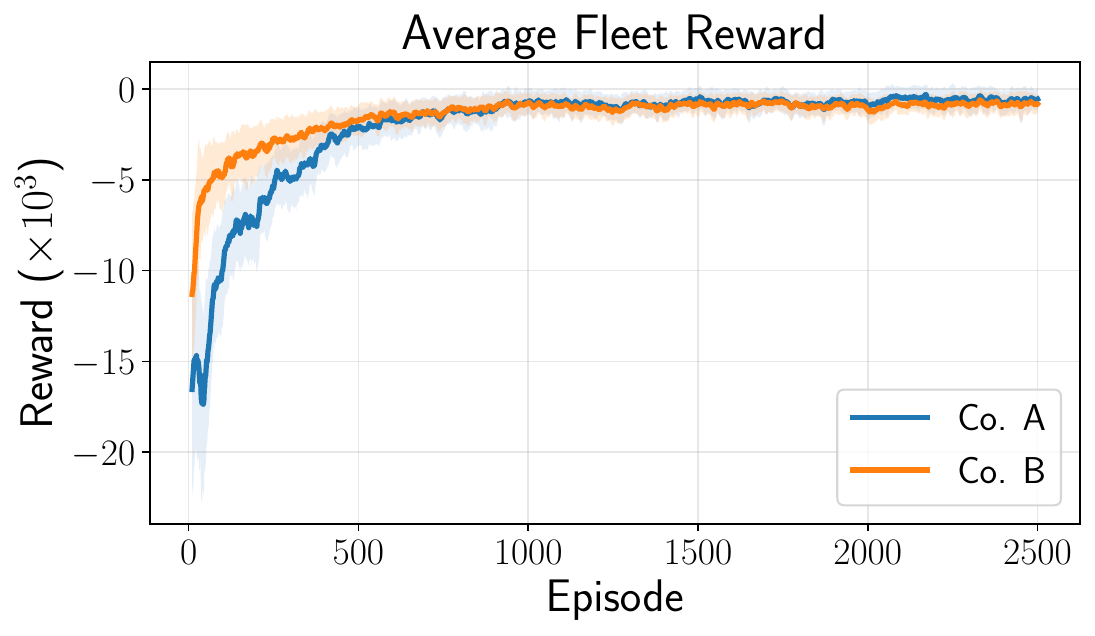}}
    \hfill
    \subfloat[Successes and NMACs]{\includegraphics[width=0.48\textwidth]{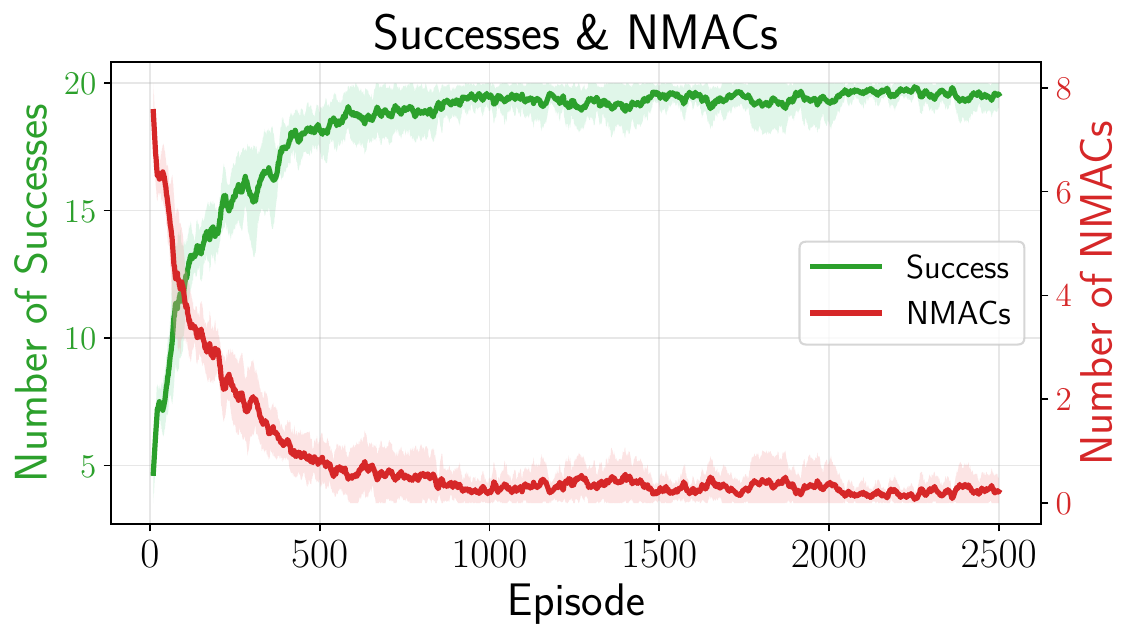}}
    \caption{Average total reward and number of successful agents with average NMACs per episode for both fleets.}
    \label{fig:rewards-success-collisions}
    \vspace{-3mm}
\end{figure*}

\subsection{\textbf{Experimental and Numerical Analyses}}

The PPOA2C framework is applied to the aforementioned complex use-case scenario to resolve intra- and inter-fleet conflicts. 

\subsubsection{\textbf{Baselines}} 
To evaluate the performance of the PPOA2C framework, we adopt a \textit{Rule-based} policy, same as~\cite{sharifi2026fine}, that operates similarly to conventional air traffic decision-making systems. Unlike the rule-based method used by Chen et al.~\cite{chen2024integrated}, which underrepresents the true performance of such methods, we enhance the rule-based approach with three additional features to make it more human-like and well-rounded. These enhancements include: (1) considering aircraft in merging/intersecting routes, not just the same route, when making decisions; (2) incorporating the distance of both the ownship and the closest intruder from the next waypoint; and (3) accounting for the nearest following intruder when determining maneuvers. This agile and informed method serves as a strong standard baseline for comparison. Additionally, we include a weak \textit{Random} baseline to demonstrate the performance of a poorly performing policy.

\subsubsection{\textbf{Policy-Configuration Combinations}}
For a comprehensive evaluation, we consider seven different combinations of policies and sUAS configurations, each expressed in the following format: policy~A (configuration~A) $+$ policy~B (configuration~B), where A and B represent two distinct fleets of agents. Policy~A and policy~B can be either \textit{Random}, \textit{Rule-based}, or \textit{PPOA2C}, while configuration~A and configuration~B can be either \texttt{X} or \texttt{Y}. The first group of policy-configuration models considers different configurations for each fleet (\texttt{X} for fleet A and \texttt{Y} for fleet B), including: Random(\texttt{X}) $+$ Random (\texttt{Y}), Rule-based(\texttt{X}) $+$ Rule-based(\texttt{Y}), PPOA2C(\texttt{X}) $+$ PPOA2C(\texttt{Y}), and PPOA2C(\texttt{X}) $+$ Rule-based(\texttt{Y}). In contrast, the second group considers identical configurations (\texttt{X}) for both fleets, including: Rule-based(\texttt{X}) $+$ Rule-based(\texttt{X}), PPOA2C(\texttt{X}) $+$ PPOA2C(\texttt{X}), and PPOA2C(\texttt{X}) $+$ Rule-based(\texttt{X}). These diverse combinations allow us to examine the influence of policy and configuration types on the overall performance of each model. During training, the PPOA2C models are updated, while the Random and Rule-based policies remain fixed.

\subsubsection{\textbf{Training Hardware}} 
All experiments were conducted using PyTorch on an NVIDIA RTX~$3090$ graphics card. Each training process consisted of $2{,}500$ episodes (approximately $600{,}000$ time steps), with each episode terminating once all agents were either truncated or removed from the simulation environment. Each episode simulated approximately $12$ minutes of flight operations, corresponding to $240$ environment time steps. Policy weights were updated at the end of each episode. On average, each complete training process required more than seven hours to finish. To ensure the reliability of results, we trained with five different random seeds (approximately $140$ hours of total training) and report the averaged outcomes across all seeds. After training, $300$ evaluation episodes were conducted to assess the effectiveness of the learned policies.

\begin{table*}[t]
\centering
\vspace{1mm}
\caption{Evaluation results for all policy-configuration combinations. A and B represent Co. A and Co. B, and \texttt{X} and \texttt{Y} are the strong and weak configurations, respectively.}
\label{tab:combined_results}
\renewcommand{\arraystretch}{1.18}
\setlength{\tabcolsep}{6pt}
\newcommand{\NA}{\multicolumn{1}{c}{—}} 
\begin{tabular}{c c c c c c c c c}
\toprule
\multicolumn{2}{l}{} & \multicolumn{4}{c}{\textbf{\texttt{XY} Configurations}} & \multicolumn{3}{c}{\textbf{\texttt{XX} Configurations}} \\
\cmidrule(r){3-6} \cmidrule{7-9}
\multicolumn{2}{l}{\makecell{\textbf{Policy A (Config. A):} \\ \textbf{Policy B (Config. B):}}} 
& \makecell{\textbf{Random(\texttt{X})} \\ \textbf{Random(\texttt{Y})}} 
& \makecell{\textbf{Rule-based(\texttt{X})} \\ \textbf{Rule-based(\texttt{Y})}} 
& \makecell{\textbf{PPOA2C(\texttt{X})} \\ \textbf{PPOA2C(\texttt{Y})}} 
& \makecell{\textbf{PPOA2C(\texttt{X})} \\ \textbf{Rule-based(\texttt{Y})}} 
& \makecell{\textbf{Rule-based(\texttt{X})} \\ \textbf{Rule-based(\texttt{X})}} 
& \makecell{\textbf{PPOA2C(\texttt{X})} \\ \textbf{PPOA2C(\texttt{X})}} 
& \makecell{\textbf{PPOA2C(\texttt{X})} \\ \textbf{Rule-based(\texttt{X})}} \\


\midrule
\multirow{7}{*}{\makecell{\textbf{Average} \\ \textbf{NMAC}}($\downarrow$)}
  & \textbf{AA}          &  3.30 & 0.24 & 0.03 & 0.000  & 0.19 & 0.02 &  0.01 \\
  & \textbf{AB}          &  1.23 & 0.25 & 0.15 & 0.005  & 0.30 & 0.26 &  0.13 \\
  & \textbf{BB}          &  3.44 & 0.26 & 0.02 & 0.001 & 0.04 & 0.02 &  0.00 \\
  \cmidrule(lr){2-9}
  & \textbf{\texttt{M1}} &  3.92 & 0.00 & 0.00 & 0.005 & 0.16 & 0.13  & 0.06 \\
  & \textbf{\texttt{M2}} &  3.96 & 0.00 & 0.00 & 0.000 & 0.13 & 0.17  & 0.08 \\
  & \textbf{\texttt{IN}} &  0.10 & 0.77 & 0.22 & 0.001 & 0.24 & 0.00 & 0.00 \\
  \cmidrule(lr){2-9}
  & \textbf{Total}  &  7.99 & 0.77 & \textbf{0.22} & \textbf{0.006} & 0.53 & \textbf{0.30} & \textbf{0.14} \\
\midrule
\rowcolor{gray!15} \textbf{Success} ($\uparrow$) & \textbf{$\textbf{N}_\textbf{s}$\,/\,20} 
& 3.96$\pm$1.77  & 18.53$\pm$1.65  & \textbf{19.55}$\pm$0.98  & \textbf{19.98}$\pm$0.10  & 19.42$\pm$0.7  & \textbf{19.79}$\pm$0.60  & \textbf{19.90}$\pm$0.34 \\
\textbf{Reward} ($\uparrow$) & $\bar{\textbf{R}}\,(\times10^3)$ & $-$ & -3.15 & \textbf{-1.41} & -1.76 & -3.27 & \textbf{-0.80} & -1.87 \\
\midrule
\multirow{2}{*}{\makecell{\textbf{Mission} \\ \textbf{Time}}} 
& $\bar{\textbf{T}}_\textbf{A}$ (min) & $-$ & 5.88 & 7.23 & 6.58 & 5.05 & 7.42 & 5.46 \\
& $\bar{\textbf{T}}_\textbf{B}$ (min) & $-$ & 7.13 & 8.94 & 7.60 & 5.05 & 7.54 & 5.02 \\
\rowcolor{gray!15} \textbf{Fairness} ($\uparrow$) & $\textbf{F}_t\,(\%)$ & $-$ & 82.4 & 80.8 & 86.5 & \textbf{100} & \textbf{98.4} & 91.9 \\
\midrule
\bottomrule
\end{tabular}
\vspace{-3mm}
\end{table*}

\subsubsection{\textbf{Training Analysis}}
Due to the large number of policy-configuration models, we only present the average total reward for models with learnable policies. We then focus on visualizing and analyzing the results of the PPOA2C(\texttt{X})$+$PPOA2C(\texttt{Y}) model. During the evaluations and comparisons, the remaining models are also considered.

\textit{\textbf{Reward Analysis}}: 
Fig.~\ref{fig:avg-tot-rewards} depicts the average total reward obtained by both fleets for the trainable models. All average rewards consistently increased and converged to small negative values--representing approximate maxima. Notably, the PPOA2C(\texttt{X})$+$PPOA2C(\texttt{Y}) and PPOA2C(\texttt{X})$+$PPOA2C(\texttt{X}) models converged to higher final values compared to the PPOA2C(\texttt{X})$+$Rule-based(\texttt{Y}) and PPOA2C(\texttt{X})$+$Rule-based(\texttt{X}) models, indicating that two heterogeneous PPOA2C policies outperformed the configurations combining one PPOA2C policy with a Rule-based policy. 

To examine the reward behavior in more detail, we focus on the most relevant model: PPOA2C(\texttt{X})$+$PPOA2C(\texttt{Y}). Fig.~\ref{fig:rewards-success-collisions} illustrates the average instantaneous rewards obtained by Co.~A and Co.~B agents for this model. Both average rewards converged to small negative values—representing approximate maxima—after about $1{,}200$ training episodes. According to the defined reward function, the maximum value is expected to be near zero, confirming that the rewards converged close to the optimal point. This outcome suggests that the learned policies not only reduce the number of NMACs but also enable smooth and efficient flight behavior consistent with the reward design. Moreover, as shown in Fig.~\ref{fig:rewards-success-collisions}, the number of successful agents ($N_{\text{s}}$), which finished their missions without NMACs, steadily increases during training and eventually converges to the maximum possible value of $20$, consistent with the observed reward trends. 

\textit{\textbf{NMAC Analysis}}:
As shown in Fig.~\ref{fig:rewards-success-collisions}, the number of NMACs ($N_{\text{c}}$) per episode decreases toward zero for the PPOA2C(\texttt{X})$+$PPOA2C(\texttt{Y}) model, although occasional NMACs still occur. We hypothesize that these rare collisions arise under highly dense traffic conditions, where agents’ actions become interdependent, and one agent’s maneuver may inadvertently cause confusion or delayed reactions in others. 

\subsection{\textbf{Evaluation Analysis and Comparison}}

During the evaluation process, the average numbers of NMACs ($N_c$) and successful missions ($N_s$) in each category were recorded for all policy-configuration models, as shown in Table~\ref{tab:combined_results}. The PPOA2C(\texttt{X})$+$PPOA2C(\texttt{Y}) and PPOA2C(\texttt{X})$+$PPOA2C(\texttt{X}) models achieved mission success rates of $97.75\%$ and $98.95\%$, respectively (an overall average of $98.35\%$). For the \texttt{XY} (\texttt{XX}) configuration settings, the PPOA2C(\texttt{X})$+$PPOA2C(\texttt{Y}) \big(PPOA2C(\texttt{X})$+$PPOA2C(\texttt{X})\big) model increased $N_s$ by $5.5\%$ ($1.9\%$) compared to the Rule-based(\texttt{X})$+$Rule-based(\texttt{Y}) \big(Rule-based(\texttt{X}) $+$ Rule-based(\texttt{X})\big) model, while reducing the average number of NMACs by $71.4\%$ ($43.3\%$), respectively. 

To further analyze NMAC occurrences, we examine them from two perspectives: (1) company-to-company (C2C) NMACs and (2) bottleneck NMACs. C2C NMACs are categorized into three types: AA, BB, and AB, representing collisions between two Co.~A agents, two Co.~B agents, and one Co.~A with one Co.~B agent, respectively. Similarly, bottleneck NMACs are divided into three categories: \texttt{M1}, \texttt{M2}, and \texttt{IN}, corresponding to the merging points \texttt{WP3} and \texttt{WP7}, and the intersection point \texttt{WP9}, respectively. Table~\ref{tab:combined_results} summarizes the C2C and bottleneck NMACs for all policy-configuration models. For the PPOA2C(\texttt{X})$+$PPOA2C(\texttt{Y}) model, approximately $75\%$ of the remaining NMACs fall under the AB category, while AA and BB NMACs are rare. NMACs still occur across all bottlenecks, but most unresolved conflicts are concentrated at \texttt{M1} and \texttt{M2}. This occurs because many aircraft fail to reach the intersection, as NMACs typically arise earlier at the merging points. Over time, \texttt{IN} NMACs disappear, whereas occasional \texttt{M1} and \texttt{M2} NMACs persist.

Across nearly all NMAC categories, the PPOA2C(\texttt{X})$+$ PPOA2C(\texttt{Y}) \big(PPOA2C(\texttt{X})$+$PPOA2C(\texttt{X})\big) model outperformed the Rule-based(\texttt{X})$+$Rule-based(\texttt{Y}) \big(Rule-based(\texttt{X}) $+$ Rule-based(\texttt{X})\big) model. This result indicates that two heterogeneous PPOA2C policies cooperate more effectively and safely than two rule-based policies in this scenario, regardless of whether the agents use different or identical configurations.

Notably, the PPOA2C policy and the Rule-based method interact even more safely than when two PPOA2C policies interact with each other. Based on Table~\ref{tab:combined_results}, the PPOA2C(\texttt{X}) $+$ Rule-based(\texttt{Y}) \big(PPOA2C(\texttt{X}) $+$ Rule-based(\texttt{X})\big) model increased $N_s$ by $2.19\%$ ($0.55\%$) compared to the PPOA2C(\texttt{X}) $+$ PPOA2C(\texttt{Y}) \big(PPOA2C(\texttt{X}) $+$ PPOA2C(\texttt{X})\big) model. However, the average evaluation reward $\bar{R}$ for the PPOA2C(\texttt{X})$+$PPOA2C(\texttt{Y}) \big(PPOA2C(\texttt{X})$+$PPOA2C(\texttt{X})\big) model is higher than that of the PPOA2C(\texttt{X})$+$Rule-based(\texttt{Y}) \big(PPOA2C(\texttt{X})$+$Rule-based(\texttt{X})\big) model. This indicates that while a PPOA2C policy and a Rule-based policy interact more safely, the interaction is not necessarily efficient. The reason lies in the Rule-based method’s tendency to adjust speed inefficiently to avoid conflicts.

\textbf{\textit{Implications:}} Based on the policy-configuration models, the results suggest that two PPOA2C policies can still learn to reach an equilibrium while maintaining near-optimal safety and efficiency. This interaction is significantly safer and more efficient than that between two well-engineered Rule-based methods. Moreover, the PPOA2C policy can also interact safely with the Rule-based method; however, such interactions are less efficient than those between two PPOA2C policies.
\vspace{-2mm}

\subsection{\textbf{Fairness Analysis}} 
When two independently trained policies interact, the learned behaviors may favor one policy over the other, potentially introducing unfair interactions. The goal here is to evaluate how the MARL framework manages this issue and whether it ensures fairness. We assess fairness based on mission time by initializing agents from both fleets under similar conditions, such as identical initial velocities. Since the travel distances are nearly equal, both fleets are expected to complete their missions in approximately the same amount of time. Given the average mission times for Co.~A and Co.~B as $\bar{T}_A$ and $\bar{T}_B$, respectively, we propose a standard time-based fairness metric, defined as $F_t(\%) = F(\bar{T}_A, \bar{T}_B)$, where $F: \mathbb{R}_{+}^2 \rightarrow [0,100]$ is expressed as:
\(F(t_1, t_2) = \left(1 - \frac{|t_1 - t_2|}{\text{max}(t_1, t_2)}\right) \times 100.\)
Higher $F_t$ values indicate greater time-based fairness between fleets. Table~\ref{tab:combined_results} presents $\bar{T}_A$, $\bar{T}_B$, and $F_t$ for all evaluation episodes. When the policy and configuration types are identical (e.g., PPOA2C(\texttt{X})$+$PPOA2C(\texttt{X})), $\bar{T}_A$ and $\bar{T}_B$ are nearly equal, resulting in a high $F_t$ of approximately $99.2\%$ on average for the corresponding policy-configuration models. However, when either the policy or configuration types differ, $\bar{T}_A$ and $\bar{T}_B$ diverge significantly, yielding lower $F_t$ values ranging from $80.8\%$ to $91.9\%$. 

When policy types are identical but configurations differ, such as in the PPOA2C(\texttt{X})$+$PPOA2C(\texttt{Y}) model, $\bar{T}_B$ is $22.4\%$ greater than $\bar{T}_A$, indicating that Co.~A, with the stronger configuration, completes the mission faster than Co.~B. This results in an average $F_t$ of $81.6\%$ for the corresponding policy-configuration models, reflecting significant bias against Co.~B with weaker configurations. Similarly, when configurations are identical but policy types differ \big(e.g., PPOA2C(\texttt{X})$+$Rule-based(\texttt{X})\big), $\bar{T}_A$ is $8.76\%$ greater than $\bar{T}_B$, with an $F_t$ of $91.9\%$. This suggests that, given equal configurations, Co.~B with a Rule-based policy operates faster than Co.~A with a PPOA2C policy. Hence, policy differences can also contribute to mission time disparities in heterogeneous settings.

\textbf{\textit{Implications:}} Based on the policy-configuration models, both policy heterogeneity and configuration heterogeneity can lead to lower time-based fairness. Furthermore, the fairness evaluation reveals that interactions consistently favor agents with stronger configurations. Even under identical configurations, differing policies may introduce discriminatory interactions among agents in terms of mission completion time.
\vspace{-2mm}

\section{Conclusion} \label{sec:conclusion}

This work investigated safe separation in dense urban airspace involving heterogeneous fleets of small unmanned aerial systems (sUASs). We employed a multi-agent reinforcement learning framework built on an attention-enhanced PPO-driven Advantage Actor-Critic (PPOA2C) algorithm and evaluated it in a realistic scenario based on the airspace over Dallas, Texas, USA. This study aimed to answer two core questions: (1) Can tactical deconfliction policies with distinct configurations reach an equilibrium to maintain a conflict-free airspace? (2) Do these policies behave fairly across fleets? Experimental results show that heterogeneous PPOA2C policies for different companies, with either similar or distinct configurations, are capable of reaching an equilibrium while outperforming strong rule-based policies in terms of mission success rate. Moreover, a PPOA2C policy demonstrates safer interaction with a Rule-based policy compared to another PPOA2C policy, although two PPOA2C policies exhibit greater efficiency based on the achieved rewards. Through extensive policy-configuration evaluations, we observed that equilibria between similar policy types exhibit bias favoring fleets with stronger configurations. Even under similar configurations but differing policy types, the equilibrium still tends to favor one of the heterogeneous policies.
\vspace{-2mm}

\bibliographystyle{ieeetr}
\bibliography{root}

\end{document}